\documentclass[prl,aps,floats,twocolumn]{revtex4-2}
\usepackage{listpackages}

\begin{document}

\title{The Big Bang as a Mirror: a Solution of the Strong CP Problem}

\author{Latham Boyle$^1$, Martin Teuscher$^{1,2}$ and Neil Turok$^{1,3}$} 
\affiliation{$^{1}$Perimeter Institute for Theoretical Physics, Waterloo, Ontario, Canada, N2L 2Y5 \\
$^{2}$\'Ecole Normale Sup\'erieure, Paris, France, 75005 \\
$^{3}$Higgs Centre for Theoretical Physics, University of Edinburgh, Edinburgh, Scotland, EH8 9YL}

\date{August 2022}

\begin{abstract}
We argue that the Big Bang can be understood as a type of mirror. We show how reflecting boundary conditions for spinors and higher spin fields are fixed by local Lorentz and gauge symmetry, and how a temporal mirror (like the Bang) differs from a spatial mirror (like the AdS boundary), providing a possible explanation for the observed pattern of left- and right-handed fermions. By regarding the Standard Model as the limit of a minimal left-right symmetric theory, we obtain a new, cosmological solution of the strong $CP$ problem, without an axion.  
\end{abstract}

\maketitle

\section{Introduction}

In a series of recent papers \cite{Boyle:2018tzc, Boyle:2018rgh, Boyle:2021jej, Boyle:2021jaz, Turok:2022fgq}, we have argued that the Big Bang can be described as a mirror separating two sheets of spacetime.  Let us briefly recap some of the observational and theoretical motivations for this idea.

Observations indicate that the early Universe was strikingly simple \cite{Planck:2015fie}: a fraction of a second after the Big Bang, the Universe was radiation-dominated, almost perfectly homogeneous, isotropic, and spatially flat; with tiny (around $10^{-5}$) deviations from perfect symmetry also taking a highly economical form: random, statistically gaussian, nearly scale-invariant, adiabatic, growing-mode density perturbations.  Although we cannot see all the way back to the bang, we have this essential observational hint: the further back we look (all the way back to a fraction of a second), the simpler and more regular the Universe gets.  This is the central clue in early Universe cosmology: the question is what it is trying to tell us. 

In the standard (inflationary) theory of the early Universe one regards this observed trend as illusory: one imagines that, if one could look back even further, one would find a messy, disordered state, requiring a period of inflation to transform it into the cosmos we observe.  An alternative approach is to take the fundamental clue at face value and imagine that, as we follow it back to the bang, the Universe really does approach the ultra-simple radiation-dominated state described above (as all observations so far {\it seem} to indicate).  Then, although we have a singularity in our past, it is extremely special \cite{penrose1979singularities}.  Denoting the conformal time by $\tau$, the scale factor $a(\tau)$ is $\propto \tau$ at small $\tau$ so the metric $g_{\mu \nu} \sim a(\tau)^2 \eta_{\mu \nu}$ has an analytic, conformal zero through which it may be extended to a ``mirror-reflected" universe at negative $\tau$ \footnote{Although it is only an approximation to treat the radiation as a perfect fluid, in the static conformal frame the fluctuations in the radiation density scale as $1/\sqrt{\cal N}$, where ${\cal N}$ is the number of relativistic degrees of freedom. Since ${\cal N}\sim 10^2$ in the Standard Model, the perfect fluid approximation is reasonable all the way to the bang.  Our mirror symmetry and analyticity conditions at the bang exclude metric perturbations that blow up there~\cite{Boyle:2018tzc, Boyle:2021jej}, and primordial black holes. Finally, including dimension zero fields which cancel the trace anomaly, the radiation fluid may be perfectly conformal near the bang~\cite{Boyle:2021jaz,perts}.}.

In \cite{Boyle:2018tzc, Boyle:2018rgh, Boyle:2021jej, Boyle:2021jaz, Turok:2022fgq} we point out that, by taking seriously the symmetries and complex analytic properties of this extended two-sheeted spacetime, we are led to elegant and testable new explanations for many of the observed features of our Universe including: (i) the dark matter \cite{Boyle:2018tzc, Boyle:2018rgh}; (ii) the absence of primordial gravitational waves, vorticity, or decaying mode density perturbations \cite{Boyle:2018tzc, Boyle:2021jej}; (iii) the thermodynamic arrow of time ({\it i.e.}\ the fact that entropy increases away from the bang) \cite{Boyle:2021jej}; and (iv) the homogeneity, isotropy and flatness of the Universe~\cite{Turok:2022fgq}, among others.  In a forthcoming paper~\cite{perts}, we show that, with our new mechanism for ensuring conformal symmetry at the bang \cite{Boyle:2021jaz}, this picture can also explain the observed primordial density perturbations. 

In this Letter, we show that: (i) there is a crucial distinction, for spinors, between spatial and temporal mirrors; (ii) the reflecting boundary conditions (b.c.'s) at the bang for spinors and higher spin fields are fixed by local Lorentz invariance and gauge invariance; (iii) they explain an observed pattern in the Standard Model (SM) relating left- and right-handed spinors; and (iv) they provide a new solution of the strong $CP$ problem~\cite{tHooft:1976rip}.

\section{Reflecting b.c.'s: spinors, higher spin}  

Locally, a mirror is a codimension-one hyperplane with a unit normal $n^{\mu}$.  Reflecting b.c.'s at such a mirror are almost uniquely fixed by local Lorentz symmetry. To describe fields of arbitrary spin, it will be convenient to work in 2-component spinor (dotted/undotted index) formalism (we follow the conventions of Appendix E in \cite{Zee:2003mt}: in particular, we will work in signature $(+,-,-,-)$\footnote{For an useful alternate introduction to 2-component spinors, see Ref.~\cite{Dreiner:2008tw}, but beware that the conventions used in this paper slightly differ from ours.

Also, the expert reader may know the peculiar fact about pinor group that $\mathrm{Pin}(1,3) \ncong \mathrm{Pin}(3,1)$, but our results are valid in both signatures $(1,3)$ and $(3,1)$ \cite{Berg:2000ne}.}). We define $\s^\mu = (I, \s^j)$, $\bar{\s}^\mu = (I, -\s^j)$ with $\s^j$ the Pauli matrices.  Under a mirror reflection, a left-chiral spinor $\varphi_{\alpha}$ is mapped to a right-chiral spinor $\bar{\chi}^{\adot}$.  Covariance under the Lorentz group implies that, at the mirror, 
\begin{subequations}
  \label{2comp_Dirac_bc}
  \begin{eqnarray}
    \label{2comp_Dirac_bc_1}
    n_{\alpha\adot}\bar{\chi}^{\adot}&=&\xi\;\!\varphi_{\alpha}\,
    \qq{where}n_{\alpha\adot}\equiv n_{\mu}(\s^{\mu})_{\alpha\adot} \\
    \label{2comp_Dirac_bc_2}
    \bar{n}^{\adot\alpha}\varphi_{\alpha}&=&\xi'\bar{\chi}^{\adot}
    \qq{where} \bar{n}^{\adot\alpha}\equiv n_{\mu}(\bar{\s}^{\mu})^{\adot\alpha}
  \end{eqnarray}
\end{subequations}
where $\xi$ and $\xi'$ are $\rm{U}(1)$-phases.  We shall call this a Dirac-type boundary condition.  Since $n_{\alpha\adot}\bar{n}^{\adot\beta}\!=\!n^{2}\delta_{\alpha}^{\beta}$ and $\bar{n}^{\adot\alpha}n_{\alpha\bdot}=n^{2}\delta^{\adot}_{\bdot}$ we see that $\xi\xi'=n^{2}$, and that Eqs.~\eqref{2comp_Dirac_bc_1} and \eqref{2comp_Dirac_bc_2} are equivalent. 

So far, $\varphi_{\alpha}$ and $\bar{\chi}^{\adot}$ are {\it independent}.  Now define the charge conjugate spinors
\begin{subequations}
  \begin{eqnarray}
    \bar{\varphi}^{\adot}\equiv\eps^{\adot\bdot}\bar{\varphi}_{\bdot} \equiv \eps^{\adot\bdot}(\varphi_{\beta}^{})^{\ast} \\ 
    \chi_{\alpha}\equiv\eps_{\alpha\beta}\chi^{\beta} \equiv \eps_{\alpha\beta}(\bar{\chi}^{\bdot})^{\ast}
  \end{eqnarray}
\end{subequations}
It is compatible with Lorentz invariance to set $\bar{\chi}^{\adot} =\bar{\varphi}^{\adot}$, {\it i.e.} so $\left(\substack{\varphi_\alpha \\ \bar{\chi}^{\adot}}\right)$ is a Majorana spinor.  With this restriction, the Dirac-type b.c. \eqref{2comp_Dirac_bc} reduces to a Majorana-type b.c.,
\begin{equation}
    \label{2comp_Majorana_bc}
    n_{\alpha\adot}\bar{\varphi}^{\adot}=\xi\varphi_{\alpha}. 
\end{equation}

Now let us generalize to higher spins. For spin 1/2, in the Dirac-type b.c.~(\ref{2comp_Dirac_bc}), we partnered every left-handed spinor $\varphi_{\alpha}$ with an independent right-handed spinor $\bar{\chi}^{\adot}$.  For higher spins, we partner $ \varphi_{\alpha_{1}\ldots\alpha_{m}}^{\bdot_{1}\ldots\bdot_{n}}$ (in the $(m/2,n/2)$ irreducible representation of the Lorentz group) with $ \bar{\chi}^{\adot_{1}\ldots\adot_{m}}_{\beta_{1}\ldots\beta_{n}}$ in the $(n/2,m/2)$ irrep.  Then the Dirac-type reflecting b.c.~\eqref{2comp_Dirac_bc} generalizes to
\begin{subequations}
  \label{general_Dirac_bc}
  \begin{eqnarray}
    n_{\alpha_{1}^{}\adot_{1}^{}}^{}\cdots\;\!n_{\alpha_{m}^{}\adot_{m}^{}}^{}
    \bar{n}^{\bdot_{1}\beta_{1}}\cdots\;\!\bar{n}^{\bdot_{n}\beta_{n}}\;
    \bar{\chi}^{\adot_{1}...\adot_{m}}_{\beta_{1}...\beta_{n}}
    &=&\xi\,\varphi_{\alpha_{1}...\alpha_{m}}^{\bdot_{1}...\bdot_{n}}\qquad \\
    \bar{n}_{}^{\adot_{1}\alpha_{1}}\!\cdots\;\!\bar{n}^{\adot_{m}\alpha_{m}}
    n_{\beta_{1}\bdot_{1}}\cdots\;\!n_{\beta_{n}\bdot_{n}}\;
    \varphi_{\alpha_{1}...\alpha_{m}}^{\bdot_{1}...\bdot_{n}}
    &=&\xi'\,\bar{\chi}^{\adot_{1}...\adot_{m}}_{\beta_{1}...\beta_{n}}\qquad
  \end{eqnarray}
\end{subequations}
where $\xi,\xi'\in\rm{U}(1)$.  Again, these are equivalent only if 
\begin{equation}
  \xi\xi'=(n^{2})^{m+n}.
\end{equation}
Likewise, we define the charge conjugate fields
\begin{subequations}
  \begin{eqnarray}
   \bar{\varphi}^{\adot_{1}...\adot_{m}}_{\beta_{1}...\beta_{n}}&=&
   \eps_{}^{\adot_{1}\dot{\g}_{1}}\cdots\;\!\eps_{}^{\adot_{m}\dot{\g}_{m}}
   \eps^{}_{\beta_{1}\delta_{1}}\cdots\;\!\eps^{}_{\beta_{n}\delta_{n}}
   \bar{\varphi}_{\dot{\g}_{1}...\dot{\g}_{m}}^{\delta_{1}...\delta_{n}}\quad \\
   \chi_{\alpha_{1}...\alpha_{m}}^{\bdot_{1}...\bdot_{n}}&=&
   \eps^{}_{\alpha_{1}\g_{1}}\cdots\;\!\eps^{}_{\alpha_{m}\g_{m}}
   \eps_{}^{\bdot_{1}\dot{\delta}_{1}}\cdots\;\!\eps_{}^{\bdot_{n}\dot{\delta}_{n}}
   \chi^{\g_{1}...\g_{m}}_{\dot{\delta}_{1}...\dot{\delta}_{n}}\quad 
  \end{eqnarray}
\end{subequations}
where $\bar{\varphi}_{\dot{\g}_{1}...\dot{\g}_{m}}^{\delta_{1}...\delta_{n}}\equiv(\varphi_{\g_{1}...\g_{m}}^{\dot{\delta}_{1}...\dot{\delta}_{n}})^{\ast}$ and $\chi^{\g_{1}...\g_{m}}_{\dot{\delta}_{1}...\dot{\delta}_{n}}\equiv(\bar{\chi}^{\dot{\g}_{1}...\dot{\g}_{m}}_{\delta_{1}...\delta_{n}})^{\ast}$ and see again that setting $\bar{\chi} =\bar{\varphi}$ is compatible with Lorentz invariance.  With this constraint, the Dirac-type b.c.~\eqref{general_Dirac_bc} reduces to the Majorana-type b.c.
\begin{equation}
    \label{general_Majorana_bc}
    \!\!n_{\alpha_{1}^{}\adot_{1}^{}}^{}\!\cdots n_{\alpha_{m}^{}\adot_{m}^{}}^{}
    \bar{n}^{\bdot_{1}\beta_{1}}\!\cdots\bar{n}^{\bdot_{n}\beta_{n}}\;
    \bar{\varphi}^{\adot_{1}...\adot_{m}}_{\beta_{1}...\beta_{n}}
    =\xi\,\varphi_{\alpha_{1}...\alpha_{m}}^{\bdot_{1}...\bdot_{n}}. \\
\end{equation}

Now, comparing Eq.~\eqref{general_Majorana_bc} to its complex conjugate and using $\eps_{\alpha\beta}\eps_{\adot\bdot}\bar{n}^{\bdot\beta} = (n_{\alpha\adot})^\ast$, we find consistency requires $(-n^{2})^{m+n}=1$. We infer that, for fermionic fields ($m+n$ odd), Majorana-type  b.c.'s are only consistent when $n^\mu$ is spacelike. The Anti de Sitter (AdS) boundary is an example: massless spinors satisfy a  Majorana-type reflecting b.c.~\eqref{general_Majorana_bc}, with a field $\varphi$ being related to its charge conjugate $\bar{\varphi}$~\cite{Avis:1977yn, Breitenlohner:1982jf, Hawking:1983mx}.   

In contrast, the Big Bang is a mirror with a timelike normal $n^\mu$. {\it The key result of this section is that fermions must satisfy a Dirac-type b.c.~\eqref{general_Dirac_bc}, where a field $\varphi$ is related to another field $\bar{\chi}$ which is not its charge conjugate.} 

\bigskip
One can check~\eqref{general_Majorana_bc} for the familiar situation of an ordinary mirror in electromagnetism. A reflection acts on spacetime as $x^{\mu}\to R^{\mu}_{\;\;\nu}x^{\nu}$, with $R^{\mu}_{\;\;\nu}=\eta^{\mu}_{\;\;\nu}-2 n^{\mu}n_{\nu}/n^{2}$. With a spacelike normal $n^\mu=(0,{\bf n})$, the b.c.'s for a perfectly conducting, ``electric" mirror are $\bm{n}\times\bm{E}=\bm{n}\cdot\bm{B}=0$. This is equivalent to imposing reflection symmetry on the field strength $F_{\kappa \lambda}=\pm R_{\kappa}^{\;\;\rho}R_{\lambda}^{\;\;\sigma}F_{\rho\sigma}$, when picking the lower sign. Then, writing $F_{\mu\nu}\to F_{\alpha\adot\beta\bdot}\to\varphi_{\alpha\beta}^{}\eps_{\dot{\alpha}\dot{\beta}}+\bar{\varphi}_{\dot{\alpha}\dot{\beta}}\eps_{\alpha\beta}^{}$, where $\bar{\varphi}$ is the self-dual part (for details see Ch. 3 and Ch. 5, p. 320 in \cite{penrose_rindler_1984}), Eq.~(\ref{general_Majorana_bc}) with $(m,n)=(2,0)$ and $\xi=1$ yields ``electric" mirror b.c.'s; while a general $\xi$ gives a mixed electric-magnetic mirror.

\section{Standard Model \& gauge invariance}

So far, our choice of b.c.'s was fixed by local Lorentz invariance. Can we make them compatible with local gauge invariance? At first glance, the answer might seem to be ``no," given that the Standard Model's chiral nature precisely means that one cannot pair up left- and right-handed spinors in this way. However, with the Higgs doublet $h$ included, the answer is  in fact ``yes." From the representation of the SM fields,~\footnote{For a pedagogical introduction to the SM, see Ref.~\cite{langacker2017standard}, particularly Section 8.1.}
\begin{equation}
  \begin{array}{c|c|c|c|c}
  && \rm{SU}(3)_C & \rm{SU}(2)_L & \rm{U}(1)_Y \\
  \hline
 \multirow{3}*{Quarks}& q_L=\left(\substack{u_L \\ d_L}\right) & 3 & 2 & +1/6 \\ 
  \cline{2-5}
  &u_R & 3 & 1 & +2/3 \\
    \cline{2-5}
 & d_R & 3 & 1 & -1/3 \\
  \hline
 \multirow{3}*{Leptons} &l_L=\left(\substack{\nu_L \\ e_L}\right) & 1 & 2 & -1/2 \\
 \cline{2-5}
 & \nu_R & 1 & 1 & 0 \\
  \cline{2-5}
 & e_R & 1 & 1 & -1 \\
  \hline
 \multirow{1}*{Higgs} & h & 1 & 2 & +1/2 \\
  \end{array}
\end{equation}
if we define  $h'=i\sigma^{2}h^{\ast}$, $\hat{h}=h/|h|$, $\hat{h}'=h'/|h'|$ and  
\begin{subequations}
  \label{uLdLnuLeL}
  \begin{eqnarray}
    u_L\equiv(\hat{h}'{}^\dagger\, q_L)&\qq{}& 
    d_L\equiv(\hat{h}^\dagger\, q_L) \\
    \nu_L\equiv(\hat{h}'{}^\dagger\;l_{L\;\!})&\qq{}&
    e_L\equiv(\hat{h}^\dagger\;l_L),
  \end{eqnarray}
\end{subequations}
it follows that $\{u_{L},d_{L},\nu_{L},e_{L}\}$ transform under $\rm{SU}(3)_C\times \rm{SU}(2)_L\times \rm{U}(1)_Y$ exactly like $\{u_{R},d_{R},\nu_{R},e_{R}\}$.  Therefore, the Standard Model's gauge symmetry is compatible with these Dirac-type boundary conditions:
\begin{subequations}
  \label{SM_bcs}
  \begin{eqnarray}
    \xi\:\! u_{L,\alpha}=n_{\alpha\adot}u_{R}^{\adot}&\qq{}&
    \xi\:\! d_{L,\alpha}=n_{\alpha\adot}d_{R}^{\adot} \\
    \xi\:\! \nu_{L,\alpha}=n_{\alpha\adot\;\!}\nu_{R}^{\adot}&\qq{}&
    \xi\:\! e_{L,\alpha}=n_{\alpha\adot}e_{R}^{\adot}
  \end{eqnarray}
\end{subequations}
with $n^{\mu}=(1,{\bf 0})$ for the bang and we can adjust the relative phase of $u_{L}$ and $u_{R}$, etc, to set $\xi=1$.

Note that $\hat{h}$ and $\hat{h}'$ live on the unit 3-sphere $\mathbb{S}^{3}$. In three spatial dimensions they are generically well-defined except on a set of measure zero, even at the bang, where $h$ satisfies a Neumann boundary condition (see below).  

This section has two main conclusions. First, for Standard Model fermions (including right-handed neutrinos), reflecting b.c.'s at the bang  \eqref{SM_bcs}  are uniquely determined by local Lorentz and gauge symmetry.  Second, reflecting  b.c's {\it require} that all Standard Model fermions can -- using the Higgs as in (\ref{uLdLnuLeL}) -- be grouped into left- and right-handed pairs that transform identically under gauge transformations. Thus, the big-bang-as-mirror hypothesis gives a new explanation for this observed fact.

\section{\texorpdfstring{Left-right symmetry and strong $\bm{CP}$}{}}  

Now consider the minimal left-right symmetric extension of the Standard Model: the LRSM. It is based on the gauge group $\rm{SU}(3)_C\times \rm{SU}(2)_L\times \rm{SU}(2)_R\times \rm{U}(1)_{B-L}$ \cite{Hall:2018let}. This theory, where each field has a left/right partner that gauge transforms similarly, has a simpler table:
\begin{equation}
  \begin{array}{r|c|c|c|c}
  & \rm{SU}(3)_C & \rm{SU}(2)_L & \rm{SU}(2)_R & \rm{U}(1)_{B-L} \\
  \hline
  q_L & 3 & 2 & 1 & +1/3 \\ 
  \hline
  q_R & 3 & 1 & 2 & +1/3 \\
  \hline
  h_L, l_L & 1 & 2 & 1 & -1 \\
  \hline
  h_R, l_R & 1 & 1 & 2 & -1 \\
  \end{array}
  \nonumber
\end{equation}
Here $h_L$ is the usual $\rm{SU}(2)_L$ Higgs doublet (previously called $h'$ in the SM) and $h_R$ is its new $\rm{SU}(2)_R$ counterpart.   If the latter acquires a vacuum expectation value, it breaks  $\rm{SU}(2)_R\times \rm{U}(1)_{B-L}$ down to $\rm{U}(1)_Y$ and the LRSM reduces to the SM below this scale.

Hall and Harigaya \cite{Hall:2018let} have argued that the LRSM is not only phenomenologically viable but has several explanatory advantages over the SM.  An independent argument~\cite{Boyle:2020ctr} is that incorporating the SM fermions into the recently-noticed connection between the SM and a special mathematical object (the exceptional Jordan algebra)~\cite{Todorov:2018mwd, Dubois-Violette:2018wgs, Baezpost}, requires embedding the SM in the LRSM.  

In the LRSM, the mirror b.c.'s at the bang take a more symmetrical form, as we can define 
\begin{subequations}
  \begin{eqnarray}
    u_{L,R}\equiv \hat{h}_{L,R}^{\dagger}\,q_{L,R}&\qq{}&  
    d_{L,R}\equiv \hat{h}_{L,R}'{}^{\!\!\!\!\!\!\!\!\dagger}\,\;\;\,q_{L,R} \\
    \nu_{L,R}\equiv \hat{h}_{L,R}^{\dagger}\;l_{L,R}&\qq{}& 
    e_{L,R}\equiv \hat{h}_{L,R}'{}^{\!\!\!\!\!\!\!\!\dagger}\;\;\;\, l_{L,R}
  \end{eqnarray}
\end{subequations}
and then write the b.c.'s as in \eqref{SM_bcs}.

These mirror b.c.'s will only yield genuine mirror symmetry between the two sheets of spacetime (on either side of the bang) if the dynamical theory is {\it also} appropriately symmetric.  We now explain the appropriate symmetry.

In our earlier paper \cite{Boyle:2018rgh}, we show in detail how $C$, $P$ and $T$ act on fields on an FRW background in which $a(\tau)$ is even or odd under $\tau\to-\tau$. As in that paper,  we use conventions where $\hbar=c=1$ and the spacetime coordinates are dimensionless so that the metric $g_{\mu\nu}$ has dimensions mass$^{-2}$; and we work for convenience in the conformal frame where the fields and couplings have all been rescaled by a power of the scale factor corresponding to their mass dimension: {\it i.e.}\ $\tilde{\varphi}(x)=a(\tau)\varphi(x)$ (for scalars), $\tilde{\psi}(x)=a^{3/2}(\tau)\psi(x)$ (for spinors), $\tilde{A}_{\mu}(x)=A_{\mu}(x)$ (for vectors); and $\tilde{g}_{\mu\nu}(x)=a^{-2}(\tau)g_{\mu\nu}(x)$ (for the metric), so the fields effectively live in a static spacetime background.  In this convention, all dimensionful couplings become functions of $\tau$; and, in particular, if $a(\tau)$ is even (resp. odd) under $\tau\to-\tau$, then couplings of odd mass dimension are even (resp. odd) under $\tau\to-\tau$.

Now consider an anti-linear $CT$ transformation which also swaps each field with its $L\leftrightarrow R$ partner, so the fields transform as:
\begin{subequations}
  \label{LR_transform}
  \begin{eqnarray}
    \label{scalar_transform}
    \tilde{h}_{L,R}(x)&\to&\tilde{h}_{R,L}(x')^{\ast}
    \qquad\qquad\quad\;\textrm{(scalars)} \\
    \label{spinor_transform}
    \tilde{\psi}_{L,R}^{i}(x)&\to&\left(\begin{array}{c} \!\gamma^{5}\! \\ 1 \end{array}\right)\gamma^{0}\tilde{\psi}_{R,L}^{i}(x')^{\ast} 
    \quad\textrm{(spinors)} \\
    \label{vector_transform}
    \tilde{A}_{\mu}^{L,R}(x)&\to&R_{\mu}^{\;\;\nu}\tilde{A}_{\nu}^{R,L}(x')^{\ast}
    \qquad\quad\;\textrm{(gauge fields)}\quad
  \end{eqnarray}
\end{subequations}
where, $R_{\mu}^{\;\;\nu}=diag(-1,1,1,1)$ is the matrix representing reflection through the bang and $x'{}^{\mu}\equiv R^{\mu}_{\;\;\nu}x^{\nu}$ is the reflected spacetime coordinate; in the spinor transformation (\ref{spinor_transform}), $\psi_{L,R}$ stands for either $q_{L,R}$ or $l_{L,R}$ and the upper/lower option ({\it i.e}\ $\gamma^{5}$ or $1$) applies when $a(\tau)$ is even/odd respectively; and in the vector transformation (\ref{vector_transform}), $A_{\mu}^{L,R}$ stands for either $W_{\mu}^{L,R}$ (the gauge fields for $\rm{SU}(2)_{L,R}$, respectively) or $G_{\mu}$ or $B_{\mu}$ (the gauge fields for $\rm{SU}(3)_{C}$ or $\rm{U}(1)_{B-L}$, which do not carry an $L/R$ label). 

Demanding the action $S_{LRSM}$ for the LRSM is invariant under this $CT$ symmetry forbids the $\theta G\tilde{G}$ dual term in the Lagrangian, because it requires that the Yukawa matrices are hermitian $Y=Y^{\dagger}$ (so there is no overall Yukawa phase, and hence the $\theta G\tilde{G}$ is not regenerated by the chiral anomaly) \footnote{See \cite{Hall:2018let} for a more detailed introduction to the LRSM Lagrangian; and for a detailed explanation of how imposing an analogous $P$ symmetry on $S_{LRSM}$ solves the strong $CP$ problem, see \cite{Babu:1989rb} and Section 4 in \cite{Hall:2018let}.  The argument is intimately related to ours, since $P$ and $CT$ symmetry are related by the $CPT$ symmetry of $S_{LRSM}$, though of course our line of reasoning, in which the bang is an actual $CT$ mirror, is physically and conceptually quite distinct.}.  (Note that demanding that $T$ rather than $CT$ symmetry would not be correct: it would eliminate the $\theta G\tilde{G}$ term, but it would also require that the Yukawa matrices are real, $Y=Y^{\ast}$, in conflict with observations.) Relatedly, note that the $S_{LRSM}$ yields classical equations that are symmetric under the corresponding linear/analytic time-reversal transformation
\begin{subequations}
  \label{LR_transform_classical}
  \begin{eqnarray}
   \label{scalarsymm}
    \tilde{h}_{L,R}(x)&\to&\tilde{h}_{R,L}(x')
    \qquad\qquad\quad\;\textrm{(scalars)} \\
    \tilde{\psi}_{L,R}^{i}(x)&\to&\left(\begin{array}{c} \!\gamma^{5}\! \\ 1 \end{array}\right)\gamma^{0}\tilde{\psi}_{R,L}^{i}(x')
    \quad\textrm{(spinors)} \\
    \tilde{A}_{\mu}^{L,R}(x)&\to&R_{\mu}^{\;\;\nu}\tilde{A}_{\nu}^{R,L}(x')
    \qquad\quad\;\textrm{(gauge fields)}\quad
  \end{eqnarray}
\end{subequations}
precisely when all the Yukawa's satisfy $Y=Y^{\dagger}$, and that a solution invariant under this transformation precisely satisfies the Dirac-like mirror boundary condition (\ref{SM_bcs}). 

In other words: {\it In the LRSM, requiring the two sheets of spacetime (before and after the bang) to be related by a mirror symmetry -- so that, at the quantum level, the bang is a surface of $CT$ symmetry (\ref{LR_transform}) and, at the classical level, the solutions of the equations of motion are invariant under the corresponding transformation (\ref{LR_transform_classical}) -- also solves the strong $CP$ problem.}

\section{Back to the Standard Model}

The left-right symmetry of the LRSM is generally broken spontaneously (see, {\it e.g.}, \cite{Hall:2018let}) because the two VEVs $\langle h_{R} \rangle$ and $\langle h_{L} \rangle$ differ in magnitude.  If $|\langle  h_{R} \rangle| \gg |\langle  h_{L} \rangle |$, at energies $\ll |\langle h_{R} \rangle|$ the LRSM reduces to the SM.  Likewise, in the limit $|\langle h_{R} \rangle|\to\infty$, the LRSM {\it is} just the SM.

How does this work in our two-sheeted, $CT$-symmetric cosmology?  
For the symmetry (13) to be gauge invariant, the gauge groups on the two sheets must be related: elements of the local gauge group $ \rm{SU}(3)_C\times \rm{SU}(2)_{L}\times \rm{SU}(2)_{R}\times \rm{U}(1)_{B-L}$ must obey $(g_{3}^{}, g_{2}^{L}, g_{2}^{R}, g_{1}^{})(\tau,x)=(g_{3}^{}, g_{2}^{R}, g_{2}^{L}, g_{1}^{})(-\tau,x)$.  This suggests that, from the two-sheeted perspective, the natural bosonic fields are $\widetilde{h}_{L,R}$ and $\widetilde{A}_{L,R}^{\mu}$, which equal $h_{L,R}$ and $A_{L,R}^{\mu}$ when $\tau>0$, and $h_{R,L}$ and $A_{R,L}^{\mu}$ when $\tau<0$.  Indeed, the limit $|\langle \widetilde{h}_{R}\rangle|\to\infty$ is compatible with our reflection symmetry, whereas $|\langle h_{R}\rangle|\to\infty$ is not. 

\begin{figure}
    \centering
    \includegraphics[scale=0.6]{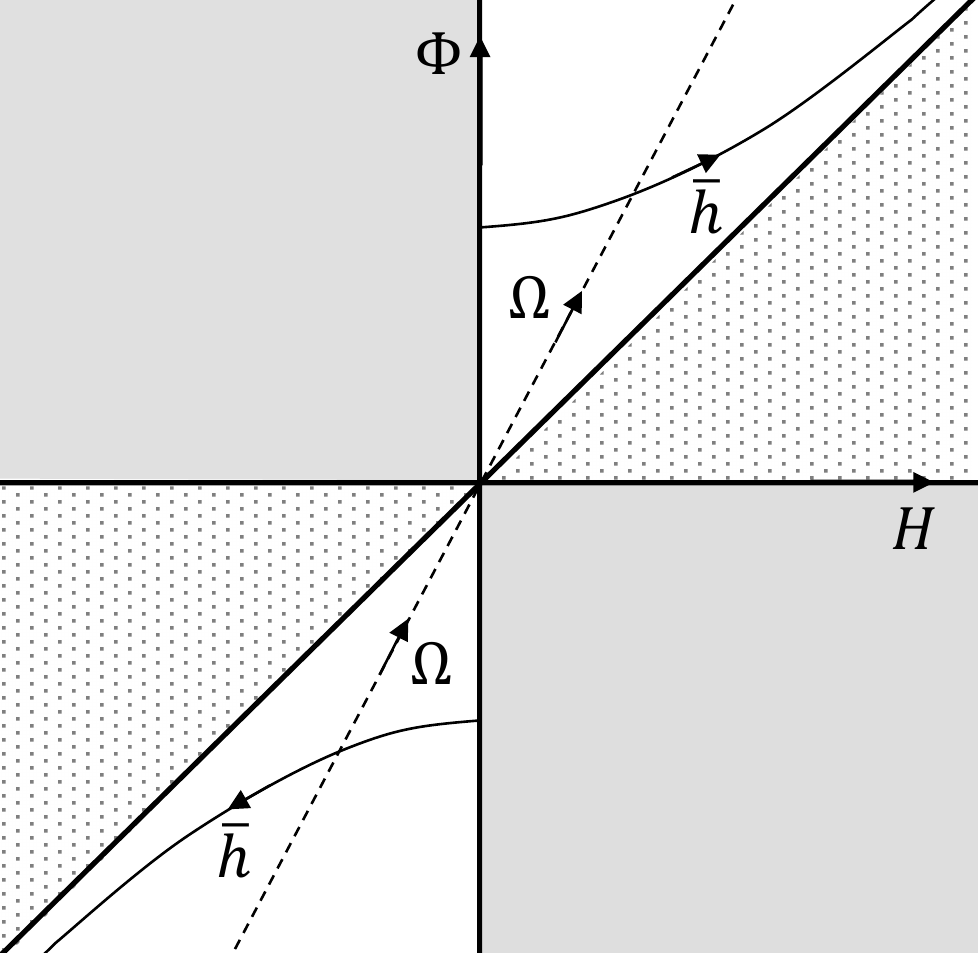}
    \caption{Cosmological solutions in the vicinity of the bang in global Minkowski $(\Phi,H)$  and local Milne $(\Omega, \bar{h})$ coordinates on field space. The solutions are straight lines in $(\Phi,H)$ (see dashed line). Mirror symmetry ensures they pass through the origin, avoiding the antigravity region (dot-filled).} 
    \label{fig:hphiplane}
\end{figure}

Further support for this view comes from coupling the theory to general relativity, and implementing a ``Weyl lift"~\cite{Bars:2013yba}. After taking the limit $|\langle \widetilde{h}_R\rangle |\to\infty$ and setting $g_{\mu \nu}=\Omega^2 \widetilde{g}_{\mu \nu}$, the Einstein-Higgs action becomes $\int  (\frac{1}{2} \Omega^2 \widetilde{R}  - 3(\partial \Omega)^2 + \Omega^2 |{\cal D}\widetilde{h}_L|^2 -\Omega^4 V(\widetilde{h}_L)) \sqrt{-\widetilde{g}}\,d^4x$ with $8\pi G=1$.  Let us work in ``unitary gauge" where
$\widetilde{h}_L\equiv\begin{psmallmatrix}\bar{h}\\0\end{psmallmatrix}$, with $\bar{h}>0$. 
Similarly, as we are interested in studying our theory on an FRW background, we can choose conformally static gauge for $\tilde{g}_{\mu\nu}$, so $\widetilde{R} =6 \kappa$ with $\kappa$ parameterizing the spatial curvature.  Near the bang the gauge field mass and spatial curvature terms vanish as $\Omega^2$ while the Higgs potential term vanishes as $\Omega^4$. Neglecting these terms, the action becomes $V_c \int ( - n^{-1}(3\,\dot{ \Omega}^2 -\Omega^2 \dot{ \bar{h}}^2) -n\, r ) d \tau,$ with $V_c$ the comoving volume, $n$ the lapse (whose variation yields the Friedmann constraint), and we have included the radiation density $r/\Omega^4$, where $r$ is a constant. Recognizing the line element on $(\Omega,\bar{h})$ field space as 2d Minkowski in Milne coordinates, we pass to global coordinates $(\Phi,H)\equiv \Omega \left(\rm{cosh}(\bar{h}/\sqrt{3}), \rm{sinh}(\bar{h}/\sqrt{3})\right)$. The action becomes $V_c \int ( -3 n^{-1}(\,\dot{ \Phi}^2 -\dot{H}^2) - n \,r ) d \tau.$  Classical trajectories consist of straight lines in the $(\Phi, H)$-plane whose 2-velocity is ``time-like" since $r>0$ (see Fig. 1). The symmetry (13) extends to $\Omega(\tau)\rightarrow -\Omega(-\tau)$, $\Phi(\tau)\rightarrow -\Phi(-\tau)$, $H(\tau)\rightarrow -H(-\tau)$: solutions satisfying this condition pass through the origin of the $(\Phi,H)$-plane and never enter the ``antigravity" region~\cite{Bars:2011aa}. As $\tau$ approaches zero, $\bar{h}$ tends to a constant, consistent with the Neumann boundary condition identified in Ref.~\cite{Boyle:2021jej}.  When finite temperature effects are included, at long wavelengths the statistical ensemble of classical saddles for $\bar{h}$ will average to zero near the bang, with $\bar{h}$ only acquiring a nonzero VEV which breaks $\widetilde{\rm{SU}(2)}_{L}\times \widetilde{\rm{U}(1)}_{Y}$ gauge symmetry at the electroweak phase transition.

Thus, although $\widetilde{h}_{R}$ corresponds to $h_{L}$ before the bang, and $h_{R}$ after it, there is no discontinuity since $h_{L}$ before the bang and $h_{R}$ after the bang parameterize the hyperbolae in two non-overlapping Milne wedges in field space, and the global coordinate $H$ is perfectly continuous. From the point-of-view of the left-right symmetric theory, the natural scalar fields are $\widetilde{h}_{L,R}$ rather than $h_{L,R}$ and the SM is then neatly recovered in the appropriate limit $|\langle\widetilde{h}_{R}\rangle|\to\infty$. Hence, our solution of the strong $CP$ problem in the LRSM extends to a corresponding solution in the SM itself! -- a subtle solution that we might never have guessed without the help of the LRSM.

\section{Classical Versus Quantum}

In this paper, we have seen how the requirement that the Big Bang is a surface of quantum $CT$ symmetry yields a new solution to the strong $CP$ problem. It also gives rise to classical solutions that are symmetric under time reversal, and satisfy appropriate reflecting boundary conditions at the bang.  The classical solutions we describe are stationary points of the action and are analytic in the conformal time $\tau$. Hence they are natural saddle points to a path integral over fields and four-geometries. The full quantum theory is presumably based on a path integral between boundary conditions at future and past infinity that are related by $CT$-symmetry.  The cosmologically relevant classical saddles inherit their analytic, time-reversal symmetry from this path integral, although the individual paths are {\it not} required to be time-symmetric in the same sense (and, moreover may, in general, be highly jagged and non-analytic).  We will describe in more detail the quantum $CT$-symmetric ensemble which implements (12), including the question of whether {\it all} of the analytic saddles are necessarily time-symmetric \cite{Newman:1992tc, Newman:1992cx}, and the calculation of the associated gravitational entanglement entropy, elsewhere~\cite{perts}. 

{\bf Acknowledgements:}  Research at Perimeter Institute is supported by the Government of Canada, through Innovation, Science and Economic Development, Canada and by the Province of Ontario through the Ministry of Research, Innovation and Science.  The work of NT is supported by the STFC Consolidated Grant `Particle Physics at the Higgs Centre' and by the Higgs Chair of Theoretical Physics at the University of Edinburgh.

\bibliography{references}

\end{document}